\documentstyle[aps,amstex,amssymb,preprint,epsf]{revtex}
\tightenlines
\begin{document}                                                       

\draft 
                                            
\title {EUCLIDEAN RESONANCE:\\
APPLICATION TO PHYSICAL AND\\ 
CHEMICAL EXPERIMENTS}

\author{B. Ivlev} 

\address
{Department of Physics and Astronomy\\
University of South Carolina, Columbia, SC 29208\\
and\\
Instituto de F\'{\i}sica, Universidad Aut\'onoma de San Luis Potos\'{\i}\\
San Luis Potos\'{\i}, S. L. P. 78000 Mexico}

\maketitle

\begin{abstract}
The phenomenon of Euclidean resonance (a strong enhancement of quantum tunneling through a nonstationary potential barrier) is applied to 
disintegration of atoms and molecules through tunnel barriers formed by applied constant and time-dependent electric fields. There are two 
different channels for such disintegration, electronic and ionic. The electronic mechanism is associated with the ionization of a molecule 
into an electron and a positive ion. The required frequencies are in a wide range between 100~MHZ and infrared. This mechanism may 
constitute a method of selective destruction of chemical bonds. The ionic mechanism consists of dissociation of a molecule into two ions. 
Since an ion is more massive than an electron, the necessary frequency is about 1~MHZ. This provides a theoretical possibility of a new 
method of isotope separation by radio frequency waves. 
\end{abstract} \vskip 0.5cm
   
 
\narrowtext

\section{INTRODUCTION}
\label{sec:intro}
Quantum tunneling through nonstationary potential barriers is very unusual. The problem was addressed in Refs.\cite{KELDYSH} and 
\cite{PERELOMOV}. The method of complex classical trajectories was developed in Refs.\cite{MELN1,MELN2,MELN3,MELN4}. Recent achievements in 
semiclassical theory are presented in Refs. \cite{KESHA,BERMAN,DEFENDI,MAITRA,ANKERHOLD,CUNIBERTI}. See also the related papers
\cite{DYAK,ZEL,GURVITZ,HANGGI}. In Refs.\cite{IVLEV1,IVLEV2} the advanced approach was developed to go beyond the method of classical 
trajectories and to obtain a space-time dependence of the wave function in semiclassical regime. 

When a nonstationary field is very small the penetration of a particle through a barrier differs hardly from a conventional tunneling 
described by the semiclassical theory of Wentzel, Kramers, and Brillouin (WKB) \cite{LANDAU}. At a relatively big nonstationary field an 
over-barrier motion enters the game when a particle should absorb a number of quanta to reach the barrier top. In this paper we consider 
soft nonstationary fields with a typical frequency much less than the barrier height. This means that a particle has to absorb a big number
of quanta to reach the barrier top. In the language of quantum mechanics, this corresponds to a high order of the perturbation theory when 
the probability is proportional to a high power of the nonstationary field. 

There are some intermediate magnitudes of a nonstationary field when neither pure tunneling nor pure over-barrier motion describes the 
penetration through a barrier. In this case, the real motion through a barrier is a combination of quanta absorption and tunneling. The 
particle pays in probability to absorb quanta and to reach the certain higher energy level but the subsequent tunneling is easier since it 
occurs in a more transparent part of the barrier. That higher energy is determined by a minimization of the total probability. This 
mechanism of barrier penetration in a nonstationary field is called photon-assisted tunneling. 

The physics of photon-assisted tunneling has no conflict with intuition since the pay in an absorption probability is compensated by the 
gain in a probability of tunneling. In photon-assisted tunneling through a triangular barrier two processes (absorption and tunneling) are 
weakly coherent which allows to consider them independently and, therefore, the total probability is a product of two partial ones. This 
reminds a static tunneling through two barriers, separated in space, when quantum coherence between them is artificially destroyed by some 
external source. In this case, two processes (tunneling and tunneling) are also independent and the total probability is a product of 
partial ones. 

Besides quanta absorption, resulting in the increase of particle energy, also quanta emission is possible followed by tunneling with a 
lower energy. At the first sight, this process cannot lead to an enhancement of tunneling due to the double loss in probability (i) 
emission of quanta and (ii) tunneling in a less transparent part of the barrier (with a lower energy). This conclusion is based on the 
assumption that quanta emission and tunneling are not strongly coherent processes which allows to consider them almost independently. The 
remarkable point of physics of nonstationary tunneling is that the processes of quanta emission and tunneling may be strongly 
{\it coherent} and cannot be considered independently. Moreover, the conclusion was drawn \cite{IVLEV2} that emission of quanta and 
tunneling with a lower energy may result in a strong enhancement of barrier penetration. This conclusion is counterintuitive. Indeed, 
intuition is sometimes useless in description of certain quantum mechanical processes. For example, the property of non-reflectivity of 
some potentials \cite{LANDAU} cannot be established on the basis of general arguments. 

The strong coherence of quanta emission and tunneling is similar to the strong coherence of two stationary tunneling processes through a 
static double barrier potential. After tunneling through the first barrier the particle performs multiple reflections from the walls of two 
barriers and then tunnels through the second barrier. Due to multiple coherent reflections two tunneling processes become strongly coherent
and the total penetration probability dramatically increases if the particle energy $E$ coincides with one of the energy levels $E_{R}$ in 
the well between two barriers. The physical idea of this mechanism, called resonant tunneling, steams to Wigner \cite{LANDAU}. The strong 
coherence between quanta emission and tunneling also results in a resonant effect and the penetration probability, as a function of a 
particle energy $E$, has a sharp peak at the certain energy $E_{R}$ determined by dynamical characteristics. This effect is called 
Euclidean resonance (ER) \cite{IVLEV2}. The difference between a stationary resonant tunneling through a static double barrier potential 
and Euclidean resonance in a dynamical barrier is that the former requires a long time for its formation but the latter occurs fast.  

One has to emphasize that the frequency of a time-dependent field is much smaller than the barrier height and the amplitude of a 
time-dependent field is smaller that the static electric field. Under these conditions the conventional tunneling and over-barrier 
transition would have very small probabilities which can be even classically small as, say, $\exp(-1000)$. In contrast to this, under the 
ER condition, $E=E_{R}$, the peak in the energy dependence of probability becomes not classically small. 

The phenomenon of Euclidean resonance can occur in any physical process where tunneling is a substantial part. In particular, ER can be 
relevant in electron emission from materials. Another example is alpha decay of nuclei which occur due to tunneling of alpha particles 
through the  Coulomb potential barrier of nuclei \cite{WEISSKOPF}. An incident proton, colliding uranium nucleus, can stimulate alpha decay
by its nonstationary Coulomb field which leads to an energy exchange between the proton and the alpha particle \cite{IVLEV3}. This 
constitutes a new type of nuclear reactions. 

The present paper addresses disintegration of atoms and molecules through tunnel barriers formed by coexisting constant and time-dependent 
electric fields. There are two different channels of such disintegration, electronic and ionic. The electronic mechanism is associated with 
ionization of a molecule into an electron and a positive ion. This can occur in a wide range of frequencies of a time-dependent field 
between 100~MHZ and infrared. The ionic mechanism is due to dissociation of a molecule into two ions. Since an ion is more massive than an
electron, the necessary frequency is about of 1~MHZ. 

Amplitudes and a frequency of applied fields can be tuned to meet ER condition for the certain electron binding energy (electronic 
mechanism) or for the certain ion dissociation energy (ionic mechanism). This allows a very selective destruction of electron chemical bonds
and a molecular dissociation only with respect to a given ion. This last mechanism provides a theoretical possibility of a new method of 
isotope separation by radio frequencies. 

The approach, used in the paper, is based on classical trajectories in imaginary time. This method enables to determine a probability of 
barrier penetration only in the exponential approximation (without a preexponential factor) but the method provides a ``bypass'' of a 
complicated quantum dynamics. 
\section{TUNNELING THROUGH A STATIC BARRIER}
\label{sec:static}
\begin{figure}[p]
\begin{center}
\vspace{1.5cm}
\leavevmode
\epsfxsize=\hsize
\epsfxsize=10cm
\epsfbox{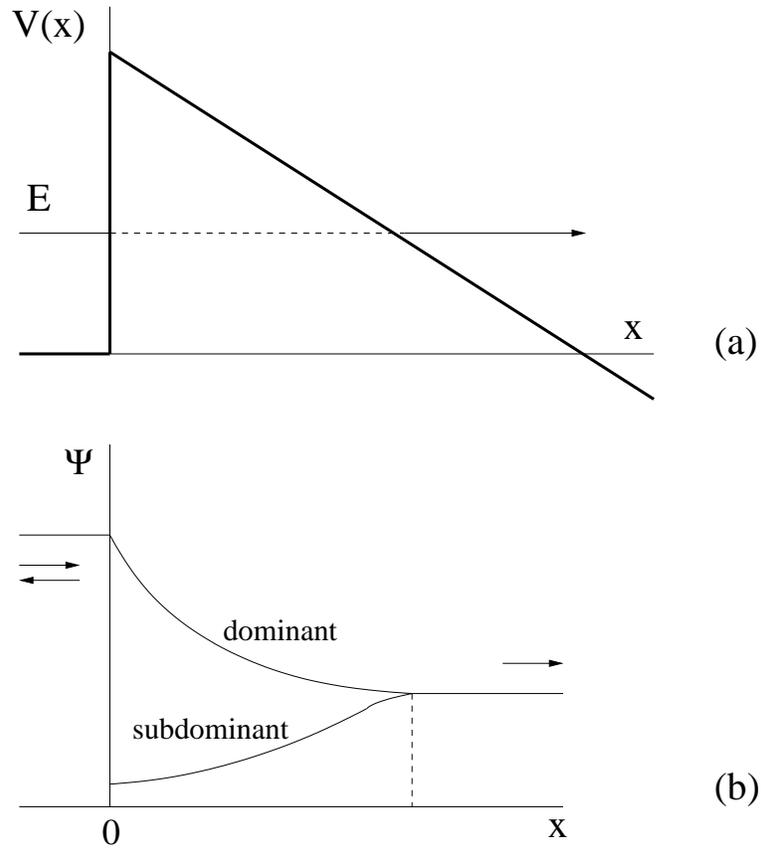}
\vspace{1cm}
\caption{(a) The electron potential barrier formed by the constant electric field. (b) The dominant branch (1) of the wave function merges 
subdominant one (2) at the classical exit point from under the barrier. There is only an outgoing wave to the right of the barrier.}
\label{fig1}
\end{center}
\end{figure}

Suppose a particle with the energy $E$ to penetrate through the static potential barrier, shown in Fig.~\ref{fig1}(a), which is 
$V-x{\cal E}_{0}$ at $x>0$ and is zero at $x<0$. If the barrier is almost classical, the flux, transmitted to the right, is exponentially
small and the particle mainly reflects from the barrier as shown by two arrows to the left from the barrier in Fig.~\ref{fig1}(b). This 
relates to damping of the wave function in a classically forbidden region corresponding to the dominant branch in Fig.~\ref{fig1}(b). To 
provide only the outgoing wave at $x\rightarrow +\infty$, shown by the arrow to the right from the barrier in Fig.~\ref{fig1}(b), the 
subdominant branch in Fig.~\ref{fig1}(b) enters the game \cite{LANDAU}. The total wave function under the barrier is a sum of two branches.
The both branches have the same order of magnitude at the exit point from under the barrier. 

The barrier penetration depends on two remarkable parameters, the barrier height $(V-E)$ and the classical time $\tau_{00}$ of an
under-barrier motion. For a weakly transparent barrier the tunneling probability is small and is given by the WKB formula \cite{LANDAU}
\begin{equation}
\label{1}
w^{(0)}\sim\exp\left[-A^{(0)}(E)\right]
\end{equation}
where
\begin{equation}
\label{1a}
A^{(0)}(E)=\frac{4}{3\hbar}(V-E)\tau_{00}
\end{equation}
The time of an under-barrier motion, for the considered triangular barrier, is 
\begin{equation}
\label{2}
\tau_{00}=\frac{\sqrt{2m(V-E)}}{{\cal E}_{0}}
\end{equation}
\section{PHOTON-ASSISTED TUNNELING}
\label{sec:phassisted}
Suppose now that the barrier is not static since its slope is modulated in time
\begin{equation}
\label{3}
V(x,t)=V-x\left({\cal E}_{0}+{\cal E}\cos\Omega t\right);\hspace{1cm}x>0
\end{equation}
and $V(x,t)=0$ at $x<0$. The amplitude ${\cal E}$ is positive. When the frequency is big, $\hbar\Omega\sim V$, the probability of the 
barrier penetration becomes not exponentially small due to over-barrier excitation of a particle \cite{KELDYSH,MELN3,MELN4}. A mostly 
interesting case corresponds to relatively small frequencies
\begin{equation}
\label{4}
\hbar\Omega\ll (V-E)
\end{equation}
which are considered in the paper.
\subsection{General approach}
According to quantum mechanical rules \cite{LANDAU}, the probability of absorption of one quantum of the field ${\cal E}\cos\Omega t$ is 
proportional to 
$\left(a{\cal E}/\hbar\Omega\right)^{2}$ where
\begin{equation}
\label{5}
a=\frac{\hbar}{\sqrt{2m(V-E)}}
\end{equation} 
is of the order of the de Broglie wave length. A particle can absorb $N$ quanta and tunnel with a higher energy where the barrier is more 
transparent (photon-assisted tunneling). The probability of this process is a product of two ones
\begin{equation}
\label{6}
w\sim\left(\frac{a{\cal E}}{\hbar\Omega}\right)^{2N}\exp\left[-A^{(0)}(E+\delta E)\right]
\end{equation}
where the energy transfer is $\delta E=N\hbar\Omega$. The simple separation (\ref{6}) of the total quantum process into two independent 
mechanisms, generally speaking, is not accurate due to quantum interference between absorption and tunneling. Nevertheless, in the case of 
triangular barrier this separation is correct \cite{MELN1,MELN2}. 

At a very small amplitude ${\cal E}$ of the nonstationary field Eq.~(\ref{6}) can be considered as a perturbative contribution to the total
penetration probability. Since $A^{(0)}(E+\hbar\Omega)\simeq A^{(0)}(E)-2\Omega\tau_{00}$, the probability of tunneling, assisted by one
photon, has the form
\begin{equation}
\label{6a}
w\sim\left[\frac{a{\cal E}}{\hbar\Omega}\exp(\Omega\tau_{00})\right]^{2}\exp\left[-A^{(0)}(E)\right]
\end{equation}
When
\begin{equation}
\label{6b}
\frac{a{\cal E}}{\hbar\Omega} >\exp(-\Omega\tau_{00})
\end{equation}
the photon-assisted tunneling (\ref{6a}) dominates the conventional one (\ref{1}) and multi-photon processes ($N>1$) have to be accounted. 
In this case, to calculate a physical probability, one should find a maximum of the expression (\ref{6}) with respect to $\delta E$. The 
maximum relates to the balance between a lose of probability due to absorption and a gain of probability due to tunneling at a higher 
energy. Proper calculations are very simple. An optimal number of absorbed quanta depends on the amplitude ${\cal E}$ 
\cite{MELN3,MELN4,IVLEV1}. The optimal energy transfer $\delta E$ increases with ${\cal E}$. Under the condition (\ref{6b}), $\delta E$ 
reaches the value $(V-E)$ and the probability takes the form
\begin{equation}
\label{7}
w\sim\exp\left[-\frac{2(V-E)}{\hbar\Omega}\ln\frac{\hbar\Omega}{a{\cal E}}\right]
\end{equation}
The expression (\ref{7}) is simply the probability to absorb $N=(V-E)/\hbar\Omega$ quanta to reach the barrier top. It is remarkable that 
at big $\Omega\tau_{00}$ tunneling processes are strongly influenced by the small amplitude (\ref{6b}) of a nonstationary field 
\cite{MELN1,MELN2,MELN3,MELN4}. 
\subsection{Quantum dynamics} 
It is very instructive to formulate photon-assisted tunneling in terms of wave function. An exact analytical solution of Schr\"{o}dinger 
equation with the nonstationary potential (\ref{3}) is impossible. Nevertheless, in the limit of small frequency (\ref{4}) one can express 
the wave function as a semiclassical series with respect to $\hbar$ 
\begin{equation}
\label{8}
\psi\left(x,t\right)=\left[a_{0}\left(x,t\right)+\hbar a_{1}\left(x,t\right)+...\right]\exp\left[\frac{i}{\hbar}S\left(x,t\right)\right].
\end{equation}
$S(x,t)$ is the classical action which mainly determines the wave function if the semiclassical conditions 
$a_{0}\gg\hbar a_{1}\gg ...\hspace{0.2cm}$ hold \cite{FEYNMAN}. This method was applied to investigation of photon-assisted tunneling in 
Ref.~\cite{IVLEV1}. The result is unusual. The semiclassical description (\ref{8}) is valid almost at all times and the wave function is of 
the type shown in Fig.~\ref{fig1}(b) for the static case. Dramatic phenomena occur close to the moments
\begin{equation}
\label{9}
t_{2n}=2n\frac{\pi}{\Omega}
\end{equation}
when the potential barrier (\ref{3}) in Fig.~\ref{fig1}(a) is mostly narrow. Close to $t=t_{2n}$ the semiclassical approach (\ref{8}) 
breaks down at the dominant branch in Fig.~\ref{fig3} within the dashed circle. During the short time interval $\hbar/V$, which cannot be 
accounted by the semiclassical approximation (\ref{8}) (instant violation of semiclassical condition), the new branch of the wave function 
is formed shown in Fig.~\ref{fig3} by the thick solid curve. Then, after an almost instant formation of the new branch, the semiclassical 
approach (\ref{8}) recovers and the new branch in Fig.~\ref{fig3} corresponds to the wave packet moving towards big $x$. This process can 
be treated as an under-barrier instability caused by a nonstationary field. With no weak quantum effects of smearing, the maximum amplitude
of the outgoing wave packet conserves and the packet moves as a classical particle. 

In the exponential approximation (without a preexponential factor), the outgoing flux of particles is determined by a maximal value of the 
new branch in Fig.~\ref{fig3}. So, the maximum of the new  branch plays an important role. The outgoing flux consists of a sequence of wave
packets, created at the moments (\ref{9}) in the barrier region, which propagate towards big $x$ as classical particles. The duration of 
each wave packet is shorter than $1/\Omega$ but longer than the de Broglie period of the order of $\hbar/V$ \cite{IVLEV1}. This enables to 
describe wave packets by the energy $E+\delta E$, where $\delta E$ is the energy transfer introduced in Sec.~\ref{sec:phassisted} A. 
Analogously to results of Ref.~\cite{IVLEV1}, the maximum of the new branch in Fig.~\ref{fig3} coincides with the value calculated on the 
basis of simple physical arguments given in Sec.~\ref{sec:phassisted} A. These simple arguments work only for truncated potential barriers 
as one given by Eq.~(\ref{3}) \cite{MELN4,IVLEV1}.

As one can see, the quantum dynamics of photon-assisted tunneling is not trivial. A violation of the semiclassical approximation (all terms
in the series (\ref{8}) become of the same order) is unavoidable during short ($\hbar/V$) intervals of time even for a slow nonstationary 
field under the condition (\ref{4}). In this situation it would be extremely useful to have some method of calculation of only a maximum of
an outgoing wave packet, avoiding complications of the full quantum dynamics. 
\subsection{Method of trajectories in imaginary time}
Let us restrict ourselves only by a calculation of a maximal value of outgoing wave packets which relates to an extreme value of the 
classical action $S$ in Eq.~(\ref{8}). According to classical mechanics, an extreme action relates to a classical trajectory. But there
are no conventional under-barrier classical trajectories since the positive value $E-V(x)=(m/2)(\partial x/\partial t)^{2}$ would be 
negative under the barrier. The famous method to avoid this problem is to use imaginary time $t=i\tau$ when 
$E-V(x)=-(m/2)(\partial x/\partial\tau)^{2}$ becomes negative. The method of complex time for photon-assisted tunneling was developed in
Refs.\cite{MELN1,MELN2,MELN3,MELN4,IVLEV1}. In our case one should find a classical trajectory $x(t_{2n}+i\tau)$ connecting the points 
$b$ and $a$ as shown in Fig.~\ref{fig3} by the dashed curve.

The classical trajectory satisfies Newton's equation 
\begin{equation}
\label{10}
m\hspace{0.1cm}\frac{\partial^{2} x}{\partial\tau^{2}}=-{\cal E}_{0}-
{\cal E}\cos\left[\Omega\left(t_{2n}+i\tau\right)\right]=-{\cal E}_{0}-{\cal E}\cosh\Omega\tau 
\end{equation}
with the boundary conditions
\begin{equation}
\label{11}
\frac{\partial x}{\partial\tau}\hspace{0.1cm}\bigg |_{\tau =\tau_{0}}=
-\sqrt{\frac{2(V-E)}{m}}\hspace{0.1cm};\hspace{1cm}x\left(t_{2n}+i\tau_{0}\right)=0
\hspace{0.1cm};\hspace{1cm}\frac{\partial x}{\partial\tau}\hspace{0.1cm}\bigg |_{\tau =0}=0
\end{equation}
The three conditions (\ref{11}) for second order differential equation (\ref{10}) define the solution and the under-barrier time $\tau_{0}$.
The moment $\tau =0$ relates to the point $b$ in Fig.~\ref{fig3}. The outgoing flux, associated with each generated wave packet, is 
proportional to the square of its amplitude and is given by the relation
\begin{equation}
\label{12}
w_{2n}\sim\exp\left(-A_{2n}\right)
\end{equation}
where the action has the form \cite{MELN3,MELN4,IVLEV1}
\begin{equation}
\label{13}
A_{2n}=\frac{2}{\hbar}\int^{\tau_{0}}_{0}d\tau\left\{\frac{m}{2}\left(\frac{\partial x}{\partial\tau}\right)^{2}+
V-x{\cal E}_{0}-x{\cal E}\cos\left[\Omega\left(t_{2n}+i\tau\right)\right]-E\right\}
\end{equation}
The last term in Eq.~(\ref{13}) corresponds to the return at $x<0$ from the time $t_{2n}+i\tau_{0}$ to $t_{2n}$ according to the 
expression $\psi\sim\exp(-iEt/\hbar)$ since at $x<0$ the nonstationary field does not act \cite{MELN3,MELN4}. So, the action $A_{2n}$ 
connects wave functions between the physical times $t_{2n}$ and $t_{2n}$. 

The solution of Eq.~(\ref{10}) should be substituted into Eq.~(\ref{13}). The field ${\cal E}\cos\Omega t$ in imaginary time goes over into 
${\cal E}\cosh\Omega\tau$ which increases exponentially and influences the trajectory only close to $\tau_{0}$. The parameter 
$\Omega\tau_{00}$ is supposed to be not small. So, the main under-barrier motion (the dashed line in Fig.~\ref{fig2}) is free with the 
energy $E+\delta E$. Only close to $\tau_{0}$ (in the vicinity of $x=0$ in Fig.~\ref{fig2}) the particle looses its energy down to $E$ 
within a short interval of imaginary time. For this reason, the process of energy transfer weakly contributes to the action (\ref{13}) 
which is mainly collected from the dashed under-barrier line in Fig.~\ref{fig2} where one can neglect the nonstationary field. In the 
expression (\ref{13}) in the value of energy $E=(E+\delta E)-\delta E$ the first part relates to the free action $A^{(0)}(E+\delta E)$ and 
$\delta E$ is an extra term. The action (\ref{13}) can be written in the form
\begin{equation}
\label{14}
A_{2n}=A^{(0)}(E+\delta E)+\frac{2}{\hbar}\hspace{0.1cm}\tau_{0}\delta E
\end{equation}
The form (\ref{14}) also follows from the exact equations (\ref{10}), (\ref{11}), and (\ref{13}) which give in the limit (\ref{6b}) the 
time of under-barrier motion
\begin{equation}
\label{14a}
\tau_{0}=\frac{1}{\Omega}\ln\frac{\hbar\Omega}{a{\cal E}}.
\end{equation}
The energy transfer $\delta E$, given by the exact equations, coincides with the value obtained from minimization of the expression 
(\ref{14}) \cite{MELN3,MELN4,IVLEV1}.

As one can see, Eqs.~(\ref{12}) and (\ref{14}) are equivalent to Eq.~(\ref{6}) followed from simple physical arguments. An importance of a 
small time-dependent field, as in Eq.~(\ref{6b}), simply follows from the imaginary time formalism where 
${\cal E}\cos(\Omega i\tau_{00})\sim {\cal E}\exp(\Omega\tau_{00})$. 
\subsection{Remarks} 
One can conclude from above that quantum dynamics of photon-assisted tunneling is complicated (instant violation of semiclassical 
condition, generation of wave packets, etc.). Nevertheless, a calculation of probability of this process in the main approximation can be 
done on the basis of the relatively simple method of trajectories in imaginary time. The trajectory provides a ``bypass'' (the dashed curve
in Fig.~\ref{fig3}) of the complicated quantum dynamics. 
\section{EUCLIDEAN RESONANCE}
\label{sec:eucl}
According to the under-barrier quantum dynamics, a creation of outgoing wave packets occurs at the moments $t=t_{2n}$ when the nonstationary
field reaches its maximal value. An amplitude of wave packets at those moments also has maximum and can be calculated by the method of 
classical trajectories as an extreme value. But besides the extremes at $t=t_{2n}$ (maxima of the field) there are also other extremes of 
the nonstationary field at $t=t_{1+2n}$ (minima of the field) where 
\begin{equation}
\label{15}
t_{1+2n}=(1+2n)\frac{\pi}{\Omega}
\end{equation}
At the moments $t_{1+2n}$ the barrier is mostly thick in accordance with Eq.~(\ref{3}). One can put a reasonable question: What happens at 
the moments $t_{1+2n}$?

Close to the moments $t_{1+2n}$ the nonstationary field (\ref{3}) is opposite to the static component ${\cal E}_{0}$ and the particle,
moving under the barrier, loses its energy. At the first sight, the particle emits (not absorbs as in Fig.~\ref{fig2}) $N$ quanta with the
probability $\left(a{\cal E}/\hbar\Omega\right)^{2N}$ and then tunnels with a lower energy. In this case the probability would be given by 
Eq.~(\ref{6}) with the negative $\delta E=-\mid\delta E\mid$ where $\mid\delta E\mid =N\hbar\Omega$. But this total probability has no 
extreme with respect to $\mid\delta E\mid$ since the both mechanisms, an emission of quanta and tunneling with a lower energy, become less 
probable under the increase of $\mid\delta E\mid$. Therefore the scenario, analogous to Fig.~\ref{fig2} but with emission instead of
absorption, is not realized. This scenario would correspond to a weak interference between tunneling and an interaction with the 
nonstationary field. As follows from above, one has to look for another mechanism related to a strong interference between quanta emission 
and tunneling.
\subsection{Quantum dynamics}
In the vicinity of the moments $t_{1+2n}$ the similar phenomenon occurs as in the vicinity of $t_{2n}$ \cite{IVLEV2}. Close to the moments 
$t_{1+2n}$ the semiclassical approximation (\ref{8}) breaks down within the dashed circle in Fig.~\ref{fig4}. In contrast to the moments 
$t_{2n}$, this happens not at the dominant branch  but at the subdominant one as shown in Fig.~\ref{fig4}. The new branch, denoted in 
Fig.~\ref{fig4} by the thick solid curve, is also formed fast during the short time interval $\hbar/V$ and then it starts to move away of 
the barrier as a wave packet. 

The maximum value of the new branch in Fig.~\ref{fig4} also plays an important role since it determines an outgoing flux with the 
exponential accuracy. But now there is no simple arguments, as at $t_{2n}$, to determine that maximum since there is no representation of
the type (\ref{6}). The maximum of the new branch in Fig.~\ref{fig4} can be calculated directly from the classical action $S$ in the 
expansion (\ref{8}) as it was done in \cite{IVLEV2} for the different shape of a nonstationary field. It is more convenient to determine 
the maximum of the new branch using the method of trajectories in imaginary time.
\subsection{Method of trajectories in imaginary time}
The imaginary time technique for the moments $t_{1+2n}$ is analogous to one developed in Sec.~\ref{sec:phassisted} C. The classical 
trajectory $x(t_{1+2n}+i\tau)$ now satisfies the equation
\begin{equation}
\label{16}
m\hspace{0.1cm}\frac{\partial^{2} x}{\partial\tau^{2}}=-{\cal E}_{0}-
{\cal E}\cos\left[\Omega\left(t_{1+2n}+i\tau\right)\right]=-{\cal E}_{0}+{\cal E}\cosh\Omega\tau
\end{equation}
with the boundary conditions (\ref{11}). The classical trajectory has to be substituted into the classical action
\begin{equation}
\label{17}
A_{1+2n}=\frac{2}{\hbar}\int^{\tau_{0}}_{0}d\tau\left\{\frac{m}{2}\left(\frac{\partial x}{\partial\tau}\right)^{2}+
V-x{\cal E}_{0}-x{\cal E}\cos\left[\Omega\left(t_{1+2n}+i\tau\right)\right]-E\right\}
\end{equation}
which is analogous to (\ref{13}) and connects wave functions between the physical times $t_{1+2n}$ and $t_{1+2n}$. The solution of equation
(\ref{16}) 
\begin{equation}
\label{18}
mx=\frac{{\cal E}_{0}}{2}\left(\tau^{2}_{0}-\tau^{2}\right)-
\frac{{\cal E}}{\Omega^{2}}\left(\cosh\Omega\tau_{0}-\cosh\Omega\tau\right)
\end{equation}
provides a connection of the points $b$ ($\tau =0$) and $a$ shown by the dashed curve in Fig.~\ref{fig4}. With the solution (\ref{18})
the action (\ref{17}) reads
\begin{equation}
\label{18a}
A_{1+2n}=\frac{{\cal E}^{2}_{0}}{m\hbar}\left[\tau_{0}\tau^{2}_{00}-\frac{\tau^{3}_{0}}{3}+
\frac{2{\cal E}}{{\cal E}_{0}}\hspace{0.1cm}\frac{\Omega\tau_{0}\cosh\Omega\tau_{0}-\sinh\Omega\tau_{0}}{\Omega^{3}}+
\left(\frac{{\cal E}}{{\cal E}_{0}}\right)^{2}\frac{2\Omega\tau_{0}-\sinh2\Omega\tau_{0}}{4\Omega^{3}}\right]
\end{equation}
The first of the boundary conditions (\ref{11}) defines the under-barrier time $\tau_{0}$ and has the form
\begin{equation}
\label{18b}
\tau_{0}-\frac{{\cal E}}{\Omega{\cal E}_{0}}\sinh\Omega\tau_{0}=\tau_{00}
\end{equation}
The outgoing flux, associated with wave packets generated at the moments $t_{1+2n}$, has the form analogous to Eq.~(\ref{12})
\begin{equation}
\label{18c}
w_{1+2n}\sim\exp\left(-A_{1+2n}\right)
\end{equation}

It is useful to look at the limit of big $\Omega\tau_{00}$. In this case the under-barrier time $\tau_{0}$, given by Eq.~(\ref{18b}), 
coincides, within the logarithmic accuracy, with the expression (\ref{14a}). Analogously to Sec.~\ref{sec:phassisted} C, at big 
$\Omega\tau_{00}$ the process of the energy transfer occurs within a short interval of imaginary time and weakly contributes to the action 
$A_{1+2n}$ which can be written, as (\ref{14}), in the form
\begin{equation}
\label{19}
A_{1+2n}=A^{(0)}\left(E-\mid\delta E\mid\right)-\frac{2}{\hbar}\hspace{0.1cm}\tau_{0}\mid\delta E\mid
\end{equation}
The energy transfer $(-\mid\delta E\mid)$ is determined by the condition $E-\mid\delta E\mid =V-{\cal E}_{0}x_{\rm exit}$ where the exit
point is $x_{\rm exit}=x(t_{1+2n})$. This $\mid\delta E\mid$ coincides with one followed from a minimization of the action (\ref{19}) 
with respect to $\mid\delta E\mid$ 
\begin{equation}
\label{20}
\tau_{0}=-\frac{\hbar}{2}\hspace{0.1cm}\frac{\partial A^{(0)}(E-\mid\delta E\mid)}{\partial E}
\end{equation}
According to classical mechanics, the right-hand side of Eq.~(\ref{20}) is a time of free under-barrier motion with the energy 
$E-\mid\delta E\mid$. The energy transfer, followed from Eq.~(\ref{20}),~is 
\begin{equation}
\label{21}
\delta E=-\left[\frac{{\cal E}^{2}_{0}}{2m\Omega^{2}}\left(\ln\frac{\hbar\Omega}{a{\cal E}}\right)^{2}-\left(V-E\right)\right]
\end{equation}
In the limit of big $\Omega\tau_{00}$, after a simple calculation within the logarithmic accuracy, one can obtain for the probability
(\ref{18c})
\begin{equation}
\label{22}
w_{1+2n}\sim\exp\left[-\frac{2(E_{R}-E)}{\hbar\Omega}\ln\frac{\hbar\Omega}{a{\cal E}}\right]
\end{equation}
In Eq.~(\ref{22}) the new remarkable energy is introduced
\begin{equation}
\label{23}
E_{R}=V-\frac{{\cal E}^{2}_{0}}{6m\Omega^{2}}\left(\ln\frac{\hbar\Omega}{a{\cal E}}\right)^{2}
\end{equation}
which is called the resonant energy. Another form of Eq.~(\ref{22}) is
\begin{equation}
\label{22a}
w_{1+2n}\sim\left(\frac{a{\cal E}}{\hbar\Omega}\right)^{2N};\hspace{1cm}N=\frac{E_{R}-E}{\hbar\Omega}
\end{equation}
Eq.~(\ref{22a}) can be interpreted as a probability to reach a top of the new reduced barrier of the height $E_{R}$. 

When the energy of the incident particle $E$ coincides with $E_{R}$, the energy transfer, as follows from Eqs.~(\ref{21}) and (\ref{23}), 
is $\delta E=-2(V-E_{R})$ and the exit point from under the barrier is determined by the relation
\begin{equation}
\label{22b}
x_{\rm exit}=\frac{3(V-E_{R})}{{\cal E}_{0}}
\end{equation}
Under the condition $E=E_{R}$ the position of the circled point in Fig.~\ref{fig4} is $(2/\sqrt{3}-1)x_{\rm exit}$. 

The probability (\ref{22}) exhibits a new feature which is absent in photon-assisted tunneling (the moments $t_{2n}$). At the energy $E$ 
close to the resonant value $E_{R}$ the probability to penetrate the barrier becomes not exponentially small. This phenomenon is called 
Euclidean resonance since the motion occurs in imaginary time. This name relates to special relativity where the metrics 
$x^{2}+y^{2}+z^{2}-(ct)^{2}$ becomes Euclidean in imaginary time $t=i\tau$. Eq.(\ref{22}) is valid at $E<E_{R}$, otherwise one has to use 
an approximation generic with a multi-instanton one. 

The resonance condition $E_{R}=E$ can be considered from another point of view, namely, when it determines the certain amplitude of the 
time-dependent field ${\cal E}_{R}(E)$ as a function of a frequency and a static field ${\cal E}_{0}$. As follows from Eqs.~(\ref{18a}) and
(\ref{18b}), the resonant value of the amplitude ${\cal E}_{R}$, defined by the condition $A_{1+2n}=0$, in the limit of big 
$\Omega\tau_{00}$ is
\begin{equation}
\label{24}
\frac{{\cal E}_{R}}{{\cal E}_{0}}=2\left(\sqrt{3}-1\right)\exp\left[-\frac{(\sqrt{3}-1)(3\sqrt{3}+1)}{4}\right]\Omega\tau_{00}
\exp\left(-\Omega\tau_{00}\sqrt{3}\right)
\end{equation}
Eqs.~(\ref{21}) - (\ref{23}) are valid for amplitudes ${\cal E}>{\cal E}_{R}$ when ${\cal E}$ is not very far from ${\cal E}_{R}$. At 
${\cal E}<{\cal E}_{R}$ a multi-instanton approximation should be used in the vicinity of the resonant value ${\cal E}_{R}$. 

With the nonstationary amplitude (\ref{24}), the probability (\ref{22}) becomes of the form 
$w_{1+2n}\sim\exp[-2\sqrt{3}(E_{R}-E)\tau_{00}/\hbar]$ that allows to estimate the resonance width in energy $\Delta E$  
\begin{equation}
\label{25}
\Delta E\sim\frac{\hbar}{2\sqrt{3}\tau_{00}}
\end{equation}
$\Delta E$ is small compared to the resonant energy $E_{R}$. 

The total outgoing flux of penetrated particles consists of wave packets created at the moments $t_{2n}$ and $t_{1+2n}$. This results in
the certain flux $W$, averaged in time, which can be experimentally measured as a steady outgoing flux. Close to the energy $E_{R}$, $W$ is
mainly determined by the wave packets created at the moments $t_{1+2n}$ and therefore, within the exponential approximation used, one can
write down $W\sim w_{1+2n}$. The probability $W$ is shown in Fig.~\ref{fig5}. 

In semiclassical approximation, $(E_{R}-E)$ cannot be less than a few of $\hbar\Omega$. Hence, the peak value of $W$ in Fig.~\ref{fig5} is
estimated, by means of Eq.~(\ref{22a}), as 
\begin{equation}
\label{26}
W\sim\left(\frac{a{\cal E}}{\hbar\Omega}\right)^{2R}
\end{equation}
where $R$ is a number of the order of unity. The above semiclassical method does not allow to calculate the number $R$ accurately. 

As one can see, the peak value of the probability (\ref{26}) is small but not semiclassically small as the expression (\ref{1}). 

According to Eq.~(\ref{24}), at big $\Omega\tau_{00}$ the nonstationary amplitude $\cal E$ is small. Therefore, it is better to choose a 
not very big $\Omega\tau_{00}$. In this case, one has to use the full equations (\ref{18a}) and (\ref{18b}). At $\Omega\tau_{00}=2$ it 
follows that $\tau_{0}\simeq 2.38\tau_{00}$ and ${\cal E}_{R}/{\cal E}_{0}\simeq 0.047$. At $\Omega\tau_{00}=1$ one can obtain 
$\tau_{0}\simeq 3\tau_{00}$ and ${\cal E}_{R}/{\cal E}_{0}\simeq 0.20$. 
\subsection{Remarks}
One dimensional tunneling through a nonstationary barrier can be formulated as a numerical problem. But performing the numerical 
calculation encounters a serious obstacle which is specific for Euclidean resonance. Since the new solution (the thick curve in 
Fig.~\ref{fig4}) is generated on the exponentially small branch, discrete steps in a numerical calculation should be chosen also 
exponentially small. Otherwise a calculation accuracy does not allow to resolve the exponentially small value, from which the instability 
develops, and the effect would be lost. The choice of extremely small steps tremendously increases a calculation time which makes the 
numerical calculation non-realistic. For a more transparent barrier this exponential value is not too small. But, in such a not very 
semiclassical case, it is unclear whether or not the above semiclassical mechanism works. 
\section{INTERPRETATION} 
\label{sec:interpret}
To understand processes underlying photon-assisted tunneling and Euclidean resonance it is useful to consider first the static problem of
tunneling through the double barrier shown in Fig.~\ref{fig6}a. The incident flux goes from the left and there is only the outgoing wave to
the right from the barriers. If the both potential barriers would be not transparent there are discrete energy levels in the well between 
them. With a finite barrier transparency each level goes over into a quasi-level with a finite width due to tunneling through the 
barriers. 
\subsection{Weak coherence}
Suppose that in the space between two barriers in Fig.~\ref{fig6}a some inelastic processes (phonons, etc.) influence the particle. This 
results in a loss of coherence between two tunneling processes across barriers which become independent. In this case the wave function 
$\psi(I)$ of the intermediate incoherent state $I$ between two barriers in Fig.~\ref{fig6}b is connected with the wave function of the 
initial state $a$ by the WKB tunneling condition $\mid\psi(I)\mid^{2}\sim\mid\psi(a)\mid^{2}\exp(-\alpha_{1})$. In the same way the state 
$b$ is linked with $I$ as $\mid\psi(b)\mid^{2}\sim\mid\psi(I)\mid^{2}\exp(-\alpha_{2})$. The total probability is
\begin{equation}
\label{27}
\frac{\mid\psi(b)\mid^{2}}{\mid\psi(a)\mid^{2}}\sim\exp\left(-\alpha_{1}-\alpha_{2}\right)
\end{equation}
This total probability to overcome the both barriers is simply a product of two partial probabilities according to the independence of the 
both processes. 

\begin{figure}[p]
\begin{center}
\vspace{1.5cm}
\leavevmode
\epsfxsize=\hsize
\epsfxsize=6cm
\epsfbox{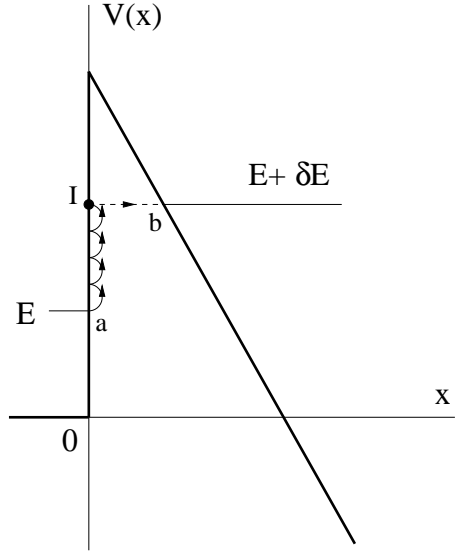}
\vspace{1cm}
\caption{The scheme of photon-assisted tunneling. A particle from the position $a$ absorbs a number of quanta, reaches the intermediate
state $I$, and then tunnels in a more transparent part of the barrier to reach the final position $b$.}
\label{fig2}
\end{center}
\end{figure}

\begin{figure}[p]
\begin{center}
\vspace{1cm}
\leavevmode
\epsfxsize=\hsize
\epsfxsize=8cm
\epsfbox{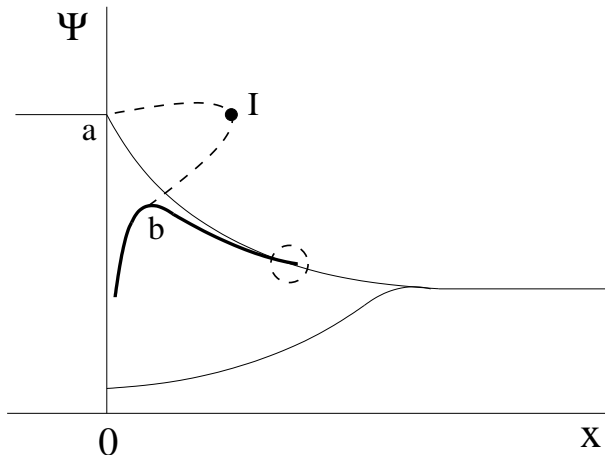}
\vspace{1cm}
\caption{Photon-assisted tunneling. The new branch, drawn by the thick curve, is formed at the circled region on the dominant branch. The 
dashed curve (``bypass'') represents the virtual process of quanta absorption $a\rightarrow I$ and tunneling $I\rightarrow b$ as in 
Fig.~\ref{fig2}.}
\label{fig3}
\end{center}
\end{figure}

\begin{figure}[p]
\begin{center}
\vspace{1.5cm}
\leavevmode
\epsfxsize=\hsize
\epsfxsize=8cm
\epsfbox{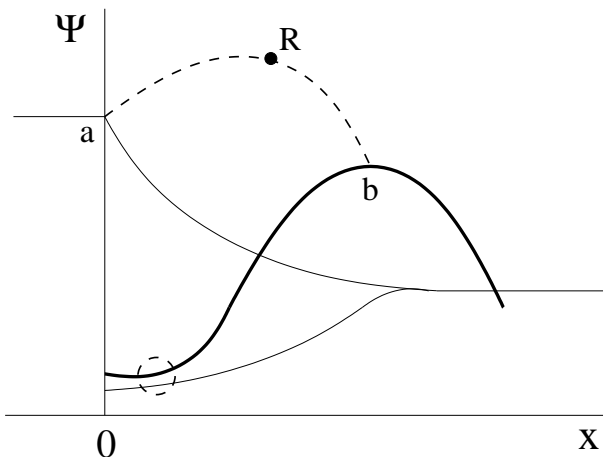}
\vspace{1cm}
\caption{Euclidean resonance. The new branch, drawn by the thick curve, is formed at the circled region on the subdominant branch. The 
dashed curve (``bypass'') represents the virtual process of quanta absorption $R\rightarrow a$ and tunneling $R\rightarrow b$ as in 
Fig.~\ref{fig7}.}
\label{fig4}
\end{center}
\end{figure}

\begin{figure}[p]
\begin{center}
\vspace{1cm}
\leavevmode
\epsfxsize=\hsize
\epsfxsize=8cm
\epsfbox{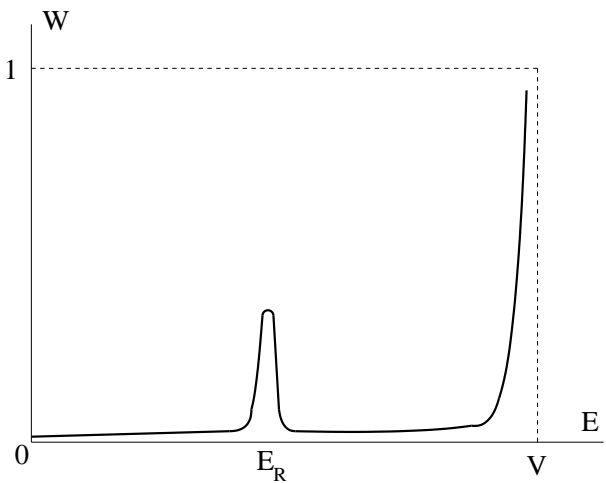}
\vspace{0.8cm}
\caption{The probability to penetrate the barrier as a function of particle energy. With a nonstationary field it has a peak at the energy 
$E_{R}$. Without a nonstationary field the probability is a monotonic function, strongly increased close to the barrier height $V$.}
\label{fig5}
\end{center}
\end{figure}

The situation is similar to photon-assisted tunneling, shown in Fig.~\ref{fig2}, where the transition $\{a\rightarrow I\}$ occurs due to 
quanta absorption (the second term in (\ref{14})) and the transition $\{I\rightarrow b\}$ relates to tunneling (the first term in 
(\ref{14})). As established in Sec.~\ref{sec:phassisted}, the both processes can be treated as incoherent ones and the total probability is
a product of two partial probabilities. The intermediate incoherent state $I$ in Fig.~\ref{fig2} corresponds to some position at the 
barrier border with the exit energy. 

Whereas the wave function for the static problem in Fig.~\ref{fig6} exists in reality, in the nonstationary case this treatment relates to 
virtual states which arise in imaginary time. The process of absorption and subsequent tunneling in Fig.~\ref{fig2} is mapped on the motion
in imaginary time shown in Fig.~\ref{fig3} by the dashed curve. The point $a$ in Figs.~\ref{fig2} and \ref{fig3} corresponds to the time 
$t_{1+2n}$ and $x=0$. The point $I$ corresponds to the time $t_{1+2n}+i\tau_{0}$ and $x=0$. The path $\{a\rightarrow I\}$ in 
Figs.~\ref{fig2} and \ref{fig3} relates to quanta absorption. The point $b$ corresponds to $t_{1+2n}$ and the path $\{I\rightarrow b\}$ 
relates to tunneling. The probability of photon-assisted tunneling is given by Eq.~(\ref{12}) where, according to the form (\ref{14}), 
$\alpha_{1}=2\tau_{0}\delta E/\hbar$ and $\alpha_{1}=A^{(0)}(E+\delta E)$. 

So, two incoherent real processes of tunneling in Fig.~\ref{fig6}b are analogous to virtual incoherent processes of absorption and tunneling
which occur in imaginary time. 
\subsection{Strong coherence}
\begin{figure}[p]
\begin{center}
\vspace{0.7cm}
\leavevmode
\epsfxsize=\hsize
\epsfxsize=13cm
\epsfbox{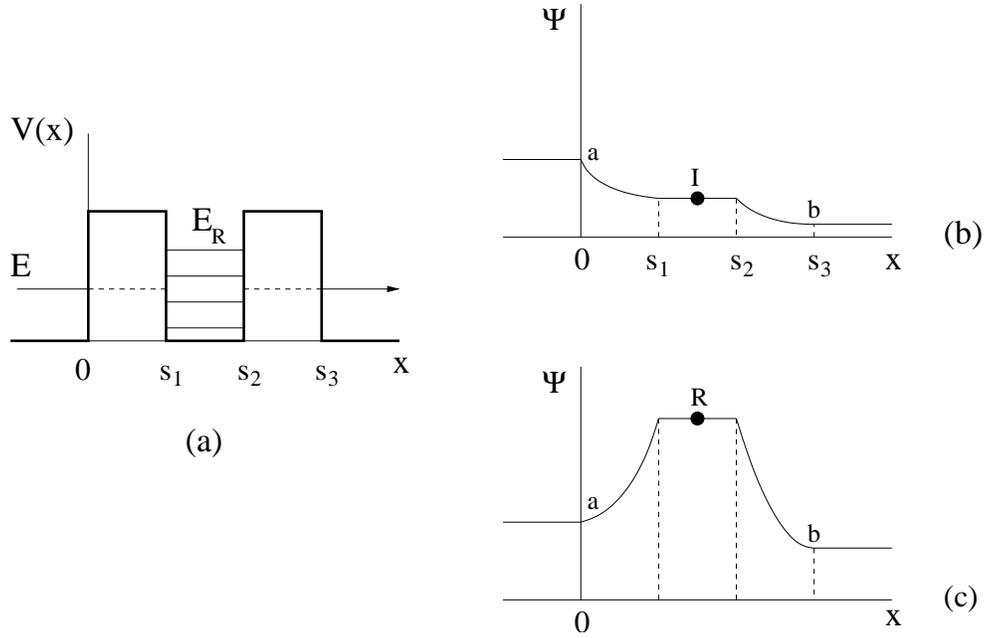}
\vspace{0.8cm}
\caption{(a) Tunneling across the double well barrier. Quasidiscrete energy levels are denoted as $E_{R}$. (b) The steady state wave 
function when the quantum coherence between two barriers is broken by some external influence. (c) The steady state wave function of the 
pure Schr\"{o}dinger problem when the natural coherence exists between two barriers.}
\label{fig6}
\end{center}
\end{figure}

When a particle motion through the static barriers in Fig.~\ref{fig6}a is not influenced by any external source, a barrier penetration is 
very peculiar. If a particle energy $E$ is close to one of the discrete levels $E_{R}$ between the barriers, the total tunneling 
probability strongly increases and can even reach unity in the case of symmetric barriers. This phenomenon is called resonant tunneling. The
idea of static resonant processes in quantum mechanics steams to Wigner \cite{LANDAU}. Two tunneling processes across the barriers
in Fig.~\ref{fig6}a become very coherent and the total probability is not reduced to a product of partial ones. 

Let us consider the wave function shown in Fig.~\ref{fig6}c. Under the resonance condition $E=E_{R}$, the static wave function in between 
the barriers is enhanced as in Fig.~\ref{fig6}c giving rise to the resonant coherent state denoted as $R$. The resonant state $R$ can be 
connected with the wave function at the barrier entrance $a$ in the form $\mid\psi(a)\mid^{2}\sim\mid\psi(R)\mid^{2}\exp(-\beta_{1})$. 
Analogously, the connection between $R$ and the exit point $b$ is $\mid\psi(b)\mid^{2}\sim\mid\psi(R)\mid^{2}\exp(-\beta_{2})$. The total 
probability of penetration the both barriers is  
\begin{equation}
\label{28}
\frac{\mid\psi(b)\mid^{2}}{\mid\psi(a)\mid^{2}}\sim\exp\left(\beta_{1}-\beta_{2}\right)
\end{equation}
Positive parameters $\beta_{1}$ and $\beta_{1}$ are determined by an exact solution of the static quantum mechanical problem. Away from the
resonance, $(\beta_{1}-\beta_{2})$ is a negative big value and the probability (\ref{27}) is exponentially small. Close to the resonance 
$E=E_{R}$ the parameter $(\beta_{1}-\beta_{2})$ reduces leading to a strong increase of the probability (\ref{28}). For symmetric barriers 
under the resonance condition $(\beta_{1}-\beta_{2})=0$, the total tunneling probability equals unity. In this case the static resonant 
phenomenon can be formulated (besides the direct way of energies coincidence $E=E_{R}$) as the equality of squared amplitudes
$\exp(-\beta_{1})=\exp(-\beta_{2})$ for the transitions $\{R\rightarrow a\}$ and $\{R\rightarrow b\}$ in Fig.~\ref{fig6}c. 

A phenomenon of Euclidean resonance has an an analogy with static resonant tunneling. In both cases the tunneling probability has a sharp
peak as a function of a particle energy at a resonant value $E_{R}$. For resonant tunneling through a static barrier, $E_{R}$ is determined
by energy levels in the potential well but for Euclidean resonance $E_{R}$ is of a dynamical origin.

\begin{figure}[p]
\begin{center}
\vspace{0.5cm}
\leavevmode
\epsfxsize=\hsize
\epsfxsize=6cm
\epsfbox{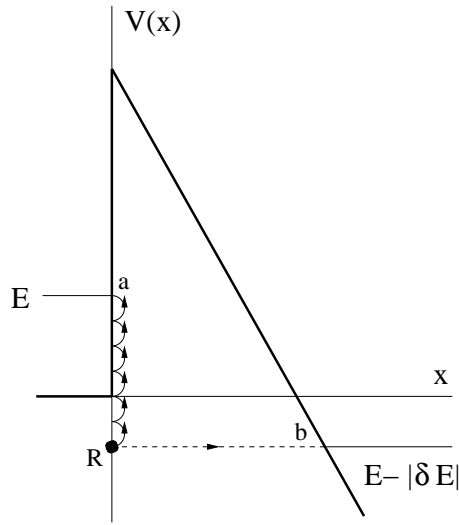}
\vspace{1cm}
\caption{The scheme of barrier penetration under conditions of Euclidean resonance. The transition $R\rightarrow a$ corresponds to 
quanta absorption and the transition $R\rightarrow b$ relates to tunneling.} 
\label{fig7}
\end{center}
\end{figure}

The coherent resonant state $R$ in Fig.~\ref{fig7} is analogous to the state $R$ between two static barriers in Fig.~\ref{fig6}c where the 
wave function is enhanced compared to the points $a$ and $b$. The resonant state $R$ in Fig.~\ref{fig7} also corresponds to an enhanced 
wave function from where one can reach the point $a$ by emission of quanta with the probability 
$\exp(-\beta_{1})\sim\exp(-2\tau_{0}\mid\delta E\mid /\hbar)$ or one can reach the point $b$ by tunneling with the probability 
$\exp(-\beta_{2})\sim\exp[-A^{(0)}(E-\mid\delta E\mid)]$. The total probability in the case of Euclidean resonance is given by 
Eq.~(\ref{28}) which coincides with the Eqs.~(\ref{18c}) and (\ref{19}). As one can see, the condition of Euclidean resonance is of the 
same type as for two static barriers, the coincidence of squared amplitudes  $\{R\rightarrow a\}$ and $\{R\rightarrow b\}$ in 
Fig.~\ref{fig7}.  

The dashed curve in Fig.~\ref{fig4} relates to the processes in Fig.~\ref{fig7} analogously to the case of photon-assisted tunneling. The 
resonant state $R$ is reached at the moment $t_{1+2n}+i\tau_{0}$ and points $a$ and $b$ correspond to the moment $t_{1+2n}$. Whereas for 
the static tunneling in Fig.~\ref{fig6}a the resonant state $R$ exists in reality, in the case of Euclidean resonance the resonance state 
$R$ is virtual and exists in imaginary time. The static resonant tunneling requires a long time for its formation (before it becomes 
steady) due to a slow process of filling in the inter-barrier space by tunneling leakage. In contrast, in a dynamical barrier the ER wave 
function is created fast. 
\section{IONIZATION OF MOLECULES}
\label{sec:ion}
The phenomenon of Euclidean resonance can provide a method of ionization of atoms and molecules into an electron and a positive ion by 
applying a constant and a time-dependent fields with properly tuned amplitudes and a frequency. The theory, developed above, can describe 
an electron tunneling from a bound state in an atom or a molecule through a potential barrier formed by applied electric fields. In the 
exponential approximation the tunneling probability is the same for both the reflection problem in Fig.~\ref{fig1} and for decay of an 
electron metastable state if it would situate at $x<0$ in Fig.~\ref{fig1} (instead of incident and reflected waves). In both cases only the
electron energy $E$ to the left of the barrier is relevant. The main physical processes occur inside the triangular barrier far from an 
initial localized bound state of a particle. For this reason, a detailed barrier shape close to a molecule (the region in the vicinity of 
$x=0$) is not very important and one can use the formalism developed above for a triangular barrier. 

As in the end of Sec.~\ref{sec:eucl}b, we choose the parameters so that $\Omega\tau_{00}=1$. In this case, according to 
Sec.~\ref{sec:eucl}b, the constant ${\cal E}_{0}$ and nonstationary ${\cal E}$ fields obey the relations (the frequency $\nu =\Omega/2\pi$ 
is measured in GHz)
\begin{equation} 
\label{29}
{\cal E}_{0}\simeq 385\nu~{\rm V}/{\rm cm};\hspace{1cm}
{\cal E}\simeq 77\nu~{\rm V}/{\rm cm};\hspace{1cm}\frac{{\cal E}}{{\cal E}_{0}}\simeq 0.20
\end{equation}
The electromagnetic energy flux $P$, the under-barrier time $\tau_{0}$, and the exit point $x_{\rm exit}$ from under the barrier are 
\begin{equation} 
\label{30}
P\simeq 15.7\nu^{2}~{\rm W}/{\rm cm}^{2};\hspace{1cm}\tau_{0}\simeq\frac{4.77}{\nu}10^{-10}~{\rm sec};\hspace{1cm}
x_{\rm exit}\simeq\frac{0.03}{\nu}~{\rm cm}
\end{equation}
The distance $x_{\rm exit}$ plays a role of the under-barrier path giving the minimal size of an experimental setup. 

The ionization rate, with the exponential accuracy (without a preexponential factor), corresponds to the expression (\ref{26}) which 
is generic with the exponent (\ref{22}). The preexponential factor can be roughly estimated as the inverse time of under-barrier motion 
$\tau^{-1}_{0}$. By means of Eq.~(\ref{30}), the ionization rate becomes of the form
\begin{equation} 
\label{31}
{\rm ionization}\hspace{0.2cm}{\rm rate}\sim\frac{(0.20)^{2R}}{\nu}10^{-10}~{\rm sec}^{-1}
\end{equation}

The conditions of Euclidean resonance in electron tunneling for different frequencies are collected in Table~\ref{tabl1}. The under-barrier
time $\tau_{0}$ and the width of Euclidean resonance $\Delta E$ (\ref{25}) are also included in Table~\ref{tabl1}. The width $\Delta E$
has only a meaning when it exceeds a thermal smearing of the electron level in the well, otherwise the thermal smearing plays a role of the
resonance width.     

Formation of Euclidean resonance can be prevented by processes of energy loss under the barrier (friction). The main friction mechanism is 
a noise of a device which produces electric fields. A small noise is not destructive. If the noise is not small it has to be of a 
relatively low frequency in order to prevent a destruction of Euclidean resonance. In this case the typical noise frequency of the electric 
field should be less than the inverse under-barrier time $\tau^{-1}_{0}$ which are listed in Table~\ref{tabl1} for each frequency. 

It is remarkable that under the above conditions, but without a nonstationary field, the tunneling probability (\ref{1}) would be 
of the order of $\exp(-1,000,000)$ as for a typical classical barrier. In the opposite case of only a nonstationary field 
(${\cal E}_{0}=0$), the probability to overcome the barrier would be determined by the process of multi-quanta absorption to reach the 
barrier top. According to Eq.~(\ref{7}), the probability of this process is also of the same type as the above classically small value. The
cooperative action of the both fields, due to the strong quantum coherence, leads to the big enhancement of barrier penetration. This is
the mostly peculiar feature of Euclidean resonance.  
\section{SELECTIVE DESTRUCTION OF CHEMICAL BONDS}
\label{sec:bonds}
In big molecules various chemical bonds have, generally speaking, different binding energies of electrons $(V-E)$. According to 
Eq.~(\ref{23}), at a fixed amplitude and a frequency of the time-dependent field, a resonant value of the constant field ${\cal E}_{0}$ 
depends on the binding energy $(V-E)$ only. This allows to adjust ${\cal E}_{0}$ to provide solely a destruction of a certain bond via 
electron tunneling with no violation of other electron bonds in the molecule. 
\section{DISSOCIATION OF MOLECULES}
\label{sec:molec}
\begin{figure}[p]
\begin{center}
\vspace{8cm}
\leavevmode
\epsfxsize=\hsize
\epsfxsize=7cm
\epsfbox{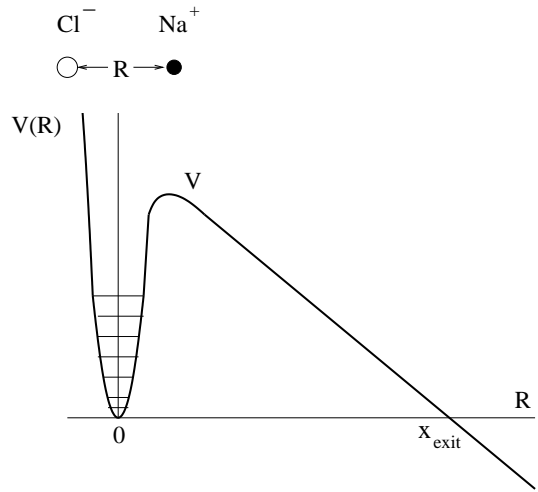}
\vspace{1cm}
\caption{The effective potential for ionization of a NaCl molecule. The vibrational levels are separated by 0.045~eV and the ionization 
energy $V$ is approximately 9~eV.}
\label{fig8}
\end{center}
\end{figure}

Tunneling mechanism may result in dissociation of molecules into ions if to put a molecule in a constant electric field as shown in 
Fig.~\ref{fig8}. If to additionally switch on a nonstationary field one can encounter conditions for Euclidean resonance. Let us consider a 
particular example of dissociation of a molecule NaCl into two ions, ${\rm Na}^{+}$ and ${\rm Cl}^{-}$ in Fig.~\ref{fig8}. The ionization 
energy, which should substitute $(V-E)$ in the electron tunneling problem, is approximately $V=9$~eV. The reduced mass of the system of two 
ions is $2.3\times 10^{-23}$ g. 

As in the previous case of electron tunneling, one can take $\Omega\tau_{00}=1$ that leads to the estimates
\begin{equation} 
\label{33}
{\cal E}\simeq 2.02\times 10^{4}\nu~{\rm V}/{\rm cm};\hspace{1cm}{\cal E}_{0}\simeq 1.01\times 10^{5}\nu~{\rm V}/{\rm cm};\hspace{1cm}
\frac{{\cal E}}{{\cal E}_{0}}\simeq 0.20
\end{equation}
The electromagnetic power flux $P$, the under-barrier time $\tau_{0}$, and the exit point $x_{\rm exit}$ from under the barrier are given
by the relations
\begin{equation} 
\label{34}
P\simeq 1.16\times 10^{6}\nu^{2}~{\rm W}/{\rm cm}^{2};\hspace{1cm}\tau_{0}\simeq\frac{4.77\times 10^{-10}}{\nu}~{\rm sec};\hspace{1cm}
x_{{\rm exit}}\simeq\frac{2.67\times 10^{-4}}{\nu}~{\rm cm}
\end{equation}
In Eqs.~(\ref{33}) and (\ref{34}) the frequency $\nu =\Omega/2\pi$ is measured in GHz. Eqs.~(\ref{33}) and (\ref{34}) allow to define
a position of Euclidean resonance for various frequencies of an external nonstationary field. For $\nu =1$~MHz (radio frequency) one can
obtain ${\cal E}_{0}\simeq 101$~V/cm, ${\cal E}\simeq 20.2$~V/cm, $P\simeq 1.16~{\rm W}/{\rm cm}^{2}$, and $x_{{\rm exit}}\simeq 0.267$~cm.

\begin{figure}[p]
\begin{center}
\vspace{1cm}
\leavevmode
\epsfxsize=\hsize
\epsfxsize=7cm
\epsfbox{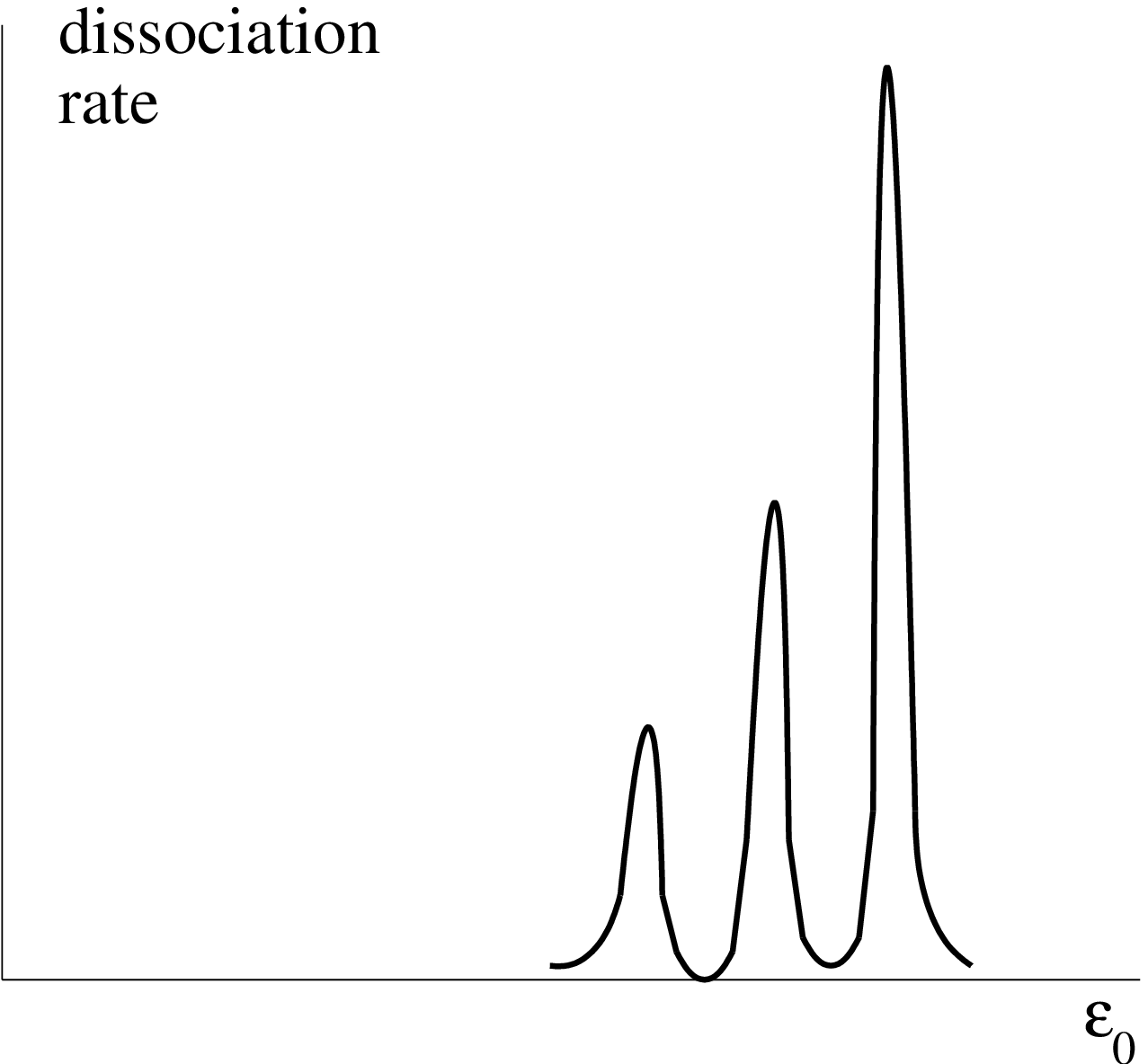}
\vspace{1cm}
\caption{Dissociation rate of a molecule as a function of an applied constant electric field under a fixed amplitude and a frequency of the 
nonstationary field. The peaks are due to thermal population of vibrational levels in Fig.~\ref{fig8}. The peak positions are sensitive to 
a molecular mass which provides a theoretical possibility of a new method of isotope separation.}
\label{fig9}
\end{center}
\end{figure}

Vibrational levels in the potential well in Fig.~\ref{fig8} for the molecule of NaCl are separated by the energy 
$\hbar\omega\simeq 0.045$~eV$\simeq$520~K. The above estimate relates to the ground state level. Suppose that an amplitude and a frequency 
of the nonstationary field are fixed and only a constant field ${\cal E}_{0}$ can vary. Then under a variation of ${\cal E}_{0}$ one can 
satisfy the condition of Euclidean resonance with respect to other energy levels corresponding to $E_{R}=\hbar\omega(1/2+n)$ in 
Eq.~(\ref{23}). This gives the positions of peaks of the dissociation rate in Fig.~\ref{fig9} 
\begin{equation} 
\label{35}
{\cal E}_{0}=\frac{\Omega\sqrt{6mV}}{\ln{\cal E}_{0}/{\cal E}}\left(1-\frac{\hbar\omega}{2V}\hspace{0.05cm}n\right);\hspace{1cm}n=0,1,2,...
\end{equation}
In our case the positions of ER peaks of the dissociation rate are
\begin{equation} 
\label{36}
{\cal E}_{0}\simeq\left(101-0.17n\right)~{\rm V}/{\rm cm}
\end{equation}
Different amplitudes of the peaks are due to different thermal occupations of the vibrational levels. 

The magnitudes of the peaks in dissociation rate, can be roughly estimated as in the analogous problem of electronic ionization in the form
$\tau^{-1}_{0}({\cal E}/{\cal E}_{0})^{2R}$. In our case it turns into    
\begin{equation} 
\label{37}
{\rm dissociation}\hspace{0.2cm}{\rm rate}\sim (0.20)^{2R}\times 10^{7}~{\rm sec}^{-1}
\end{equation}
The width of peaks $\Delta{\cal E}_{0}$ in Fig.~\ref{fig9} can be defined from the relation 
$\Delta{\cal E}_{0}/{\cal E}_{0}\sim\Delta E/V$. The energy width $\Delta E\sim 10^{-4}$~K is given by Eq.~(\ref{25}) and is much less than
the natural width $\Delta E_{T}$ of vibrational levels determined by thermal processes. Therefore, the peak width in Fig.~\ref{fig9} is
given by $\Delta{\cal E}_{0}\sim{\cal E}_{0}\Delta E_{T}/V$. 

In complicated molecules it is possible to tune field amplitudes and a frequency in order to destroy a certain vibrational molecular mode
via ionic tunneling through the barrier. This is an ionic selective destruction of molecules in addition to the electronic mechanism 
considered in Sec.~\ref{sec:ion}.

One should emphasize that the radio frequency quantum used is $10^{7}$ times smaller than the distance $\hbar\omega$ between vibrational 
levels in Fig.~\ref{fig8}. The considered effect of dissociation is not due to a redistribution of levels occupation in the potential well
in Fig.~\ref{fig8} but due to under-barrier processes in presence of the time-dependent field.

In the absence of a nonstationary field the tunneling probability is classically small as $\exp(-10^{9})$. The same can be concluded about 
the probability to absorb $V/\hbar\Omega\sim 10^{9}$ quanta to reach the barrier top in the absence of a constant field. 
\section{A NEW METHOD OF ISOTOPE SEPARATION}
\label{sec:sep}
There are various methods of isotope separation based on diffusion, centrifugal forces, electromagnetic forces, laser irradiation, and 
chemical processes. A phenomenon of Euclidean resonance adds a new theoretical possibility of isotope separation (by radio frequency waves)
to the previous list. The method is based on induced dissociation of molecules considered in Sec.\ref{sec:molec}. If the isotope mass shift
is $\Delta m$, then the relative shift of resonant values of the constant field in Fig.~\ref{fig9} is $\Delta m/2m$. When 
$\Delta m/m\simeq 0.01$ the isotope shift of ER peaks in Fig.~\ref{fig9} is approximately 1~V/cm. This enables to tune the system to ionize
only the isotope of the mass $m+\Delta m$.
\section{DISCUSSION AND CONCLUSIONS}
\label{sec:disc}
The action of a nonstationary field on quantum tunneling is very non-trivial. As soon as the nonstationary field is small, one can use a 
perturbation theory with respect to its amplitude and the total effect is simply reduced to an absorption or an emission of a few quanta 
and a subsequent tunneling. A physical scenario becomes completely different with the increase of an amplitude of the nonstationary field. 
Methods of perturbation theory break down but indicate that multiquanta processes become relevant. In this case, as shown in the paper, the
barrier penetration is realized as an emission of outgoing wave packets at the moments of time when the barrier becomes of an extremal 
width. The mostly thin barrier relates to photon-assisted tunneling and the mostly thick one corresponds to Euclidean resonance. The 
quantum dynamic of the both physical processes is similar to some extend. Due to an under-barrier instability the wave packet is quickly 
formed and then it moves out as a classical particle. 

Nevertheless, there is an essential difference between the two processes. Suppose the static barrier to be classically non-transparent with 
the tunneling probability of the type $\exp(-1000)$. Then due to photon-assisted tunneling a particle increases its energy and exits from 
under the barrier with the probability, say, $\exp(-600)$. Due to Euclidean resonance a particle looses its energy but penetrates the 
barrier with the probability given by Eq.~(\ref{26}) which is small but not classically small as the above numbers.   

The regime of Euclidean resonance is extremely unusual since a classical potential barrier can be penetrated by particles acted by a 
time-dependent electric field of a frequency which is much less compared to the barrier height. Moreover, an electron looses its energy 
and, at the first sight, seems to travel a longer way under the barrier. This contradicts to the traditional imagination of under-barrier 
motion, a quanta absorption and a subsequent tunneling, based on a weak coherence between these processes. The reason of this contradiction 
is that the phenomenon of Euclidean resonance relates to a {\it strong} coherence between electromagnetic interaction and tunneling when 
a separation of a total process into almost independent emission and tunneling is impossible.

The method of classical trajectories in imaginary time, developed in the paper, enables to determine the magnitude of the effect with no
detailed quantum mechanical calculations. This method provides a ``bypass'' of the complicated quantum dynamics including the regions 
where semiclassical approximation breaks down.   

In the paper two tunneling mechanisms, electronic and ionic, of molecule destruction are considered. The electronic destruction 
(ionization) relates to electron tunneling through a potential barrier, formed by a constant electric field, in presence of a 
time-dependent one. The ionic destruction (dissociation) is associated with the analogous process participated by an ion. An ion is more
massive than an electron and moves slower under the barrier. Therefore a smaller frequency (roughly 1~MHZ) is required for the ionic 
dissociation than for the electronic ionization (more than 100~MHZ).  

In a complicated molecule one can selectively destroy a particular electronic chemical bond since ER peak position depends on a binding 
electron energy. In addition to this, a selective ionic dissociation of molecules is possible by tuning external constant and 
time-dependent fields to a certain ionic motion. 

Since a position of ER peak depends sensitively on an ion mass this provides a theoretical possibility of a new method of isotope 
separation if to tune properly field magnitudes and a frequency. This method is unusual since radio frequencies of the order of 1~MHz have 
never been used for isotope separation.

Generally speaking, the phenomenon of Euclidean resonance is applicable to a variety of phenomena where quantum tunneling is a substantial
part of physical processes. In particular, one can consider a possibility of ER application in scanning tunneling microscopy (STM) 
\cite{STROSCIO,GOMER,GRAF}, in molecular electronics \cite{SEM}, nanoscience, for tunneling chemical reactions \cite{MIYA}, and for decay
of zero-voltage state of Josephson junctions \cite{BARONE,CLARKE,USTINOV}. On the basis of Euclidean resonance a prediction of a new type 
of nuclear reaction has been made \cite{IVLEV3}. 
\acknowledgments
I am grateful to G.~Berman, L.~Bulaevskii, J.~Engelfried, V.~Gudkov, S.~Gurvitz, M.~Kirchbach, J.~Knight, M.~Kunchur, L.~Levitov, 
R.~Prozorov, A.~Redondo, J.~Seminario, S.~Trugman, and V.~Tsifrinovich for valuable discussions.

\newpage

\begin{table}
\centering
\begin{tabular}{|c|c|c|c|}
\hline
&radio frequency&microwaves&infrared\\
\hline
frequency&$\nu$=100~MHz&$\nu$=1~GHz&$\nu =3\times 10^{4}~{\rm GHz}$\hspace{0.2cm}$(\lambda =10~\mu{\rm m})$\\
dc field (V/cm)&38.5&385&$1.15\times 10^{7}$\\
ac power $({\rm W}/{\rm cm}^{2})$&0.15&15&$1.41\times 10^{10}$\\
under-barrier path $x_{{\rm exit}}$ (cm)&0.3&0.03&$10^{-6}$\\
under-barrier time $\tau_{0}$ (sec)&$4.77\times 10^{-9}$&$4.77\times 10^{-10}$&$1.59\times 10^{-14}$\\
width $\Delta E$ (K)&0.0013&0.013&390\\
\hline
\end{tabular}
\vspace{1cm}
\caption{The conditions of Euclidean resonance in electron tunneling for different frequencies}\label{tabl1}
\end{table}

\newpage

\begin{references}

\bibitem{KELDYSH}

L.V. Keldysh, Zh. \'{E}ksp. Teor. Fiz. {\bf 47}, 1945 (1964) [Sov. Phys. JETP {\bf 20}, 1307 (1965)].

\bibitem{PERELOMOV}

V.S. Popov, V.P. Kuznetsov, and A.M. Perelomov, Zh. \'{E}ksp. Teor. Fiz. {\bf 53}, 331 (1967). 
[Sov. Phys. JETP {\bf 26}, 222 (1968)].

\bibitem{MELN1}

B.I. Ivlev and V.I. Melnikov, Pis'ma Zh. \'{E}ksp. Teor. Fiz. {\bf 41}, 116 (1985) [JETP Lett. {\bf 41}, 142 (1985)].

\bibitem{MELN2}

B.I. Ivlev and V.I. Melnikov, Phys. Rev. Lett. {\bf 55}, 1614 (1985).

\bibitem{MELN3}

B.I. Ivlev and V.I. Melnikov, Zh. \'{E}ksp. Teor. Fiz. {\bf 90}, 2208 (1986) [Sov. Phys. JETP {\bf 63}, 1295 (1986)]. 

\bibitem{MELN4}

B.I. Ivlev and V.I. Melnikov, in {\it Quantum Tunneling in Condensed Media}, edited by\\ 
A. Leggett and Yu. Kagan (North-Holland, Amsterdam, 1992).

\bibitem{KESHA}

S. Keshavamurthy and W.H. Miller, Chem. Phys. Lett. {\bf 218}, 189 (1994).

\bibitem{BERMAN}

T. Martin and G. Berman, Phys. Lett. A {\bf 196}, 65 (1994).

\bibitem{DEFENDI}

A. Defendi and M. Roncadelli, J. Phys. A {\bf 28}, L515 (1995).

\bibitem{MAITRA}

N.T. Maitra and E.J. Heller, Phys. Rev. Letter. {\bf 78}, 3035 (1997).

\bibitem{ANKERHOLD}

J. Ankerhold and H. Grabert, Europhys. Lett. {\bf 47}, 285 (1999).

\bibitem{CUNIBERTI}

G. Cuniberty, A. Fechner, M. Sassetti, and B. Kramer, Europhys. Lett. {\bf 48}, 66 (1999).

\bibitem{DYAK}

M.I. Dyakonov and I.V. Gornyi, Phys. Rev. Lett. {\bf 76}, 3542 (1996).

\bibitem{ZEL}

C.A. Bertulani, D.T. de Paula, and V.G. Zelevinsky, Phys. Rev. C {\bf 60}, 031602 (1999).

\bibitem{GURVITZ}

S.A. Gurvitz, Phys. Rev. A {\bf 38}, 1747 (1988).

\bibitem{HANGGI}

L. Hartmann, I. Gouchuk, and P. H\"{a}nggi, J. Chem. Phys. {\bf 113}, 11159 (2000).

\bibitem{IVLEV1}

B.I. Ivlev, Phys. Rev. A {\bf 62}, 062102 (2000).

\bibitem{IVLEV2}

B.I. Ivlev, Phys. Rev. A {\bf 66}, 012102 (2002).

\bibitem{LANDAU}

L.D. Landau and E.M. Lifshitz, {\it Quantum Mechanics} (Pergamon, New York, 1977).

\bibitem{WEISSKOPF}

J.M. Blatt and V.F. Weisskopf, {\it Theoretical Nuclear Physics} (Springer-Verlag, New York, 1979).

\bibitem{IVLEV3}

B. Ivlev and V. Gudkov, Phys. Rev. C {\bf 69}, 037602 (2004).

\bibitem{FEYNMAN}

R.P. Feynman and A.R. Hibbs, {\it Quantum Mechanics and Path Integrals} (McGrow-Hill, New York, 1965).

\bibitem{STROSCIO}

{\it Scanning Tunneling Microscopy}, edited by J.A. Stroscio and W.J. Kaiser (Academic Press, San Diego 1993).

\bibitem{GOMER}

{\it Field Emission and Field Ionization}, R. Gomer (American Institute of Physics, New York, 1993).

\bibitem{GRAF}

S. Grafstr\"{o}m, Journ. Appl. Phys. {\bf 91}, 1717 (2002).

\bibitem{SEM}

J.M. Seminario, A.G. Zacarias, and J.M. Tour, J. Am. Chem. Soc. {\bf 122}, 3015 (2000).

\bibitem{MIYA}

T. Miayzaki, J. Nucl. Sci. and Technology, {\bf 39}, 339 (2002).

\bibitem{BARONE}

A. Barone and G. Patern\`{o}, {\it Physics and Application of the Josephson Effect} (Wiley, New York, 1982).

\bibitem{CLARKE}

M.H. Devoret, D. Esteve, C. Urbina, J. Martiniis, A. Cleland, J. Clarke, in {\it Quantum Tunneling in Condensed Media}, edited by
A. Leggett and Yu. Kagan (North-Holland, Amsterdam, 1992).

\bibitem{USTINOV}

A. Wallraff, A. Lukashenko, J. Lisenfeld, A. Kemp, M.V. Fistul, Y. Koval, and A.V. Ustinov, Nature {\bf 425}, 155 (2003).


\end{references}
\end{document}